\documentstyle[psfig,prd,aps]{revtex}
\begin{document}

\draft
\title{The small $x$ nuclear shadowing at DIS}
\author{
  M. B. Gay  Ducati $^{*}$\footnotetext{$^{*}$E-mail:gay@if.ufrgs.br}
 and
Victor P.  Gon\c{c}alves $^{**}$\footnotetext{$^{**}$E-mail:barros@if.ufrgs.br}} 
\address{ Instituto de F\'{\i}sica, Universidade Federal
do Rio Grande do Sul\\ Caixa Postal 15051, CEP 91501-970, Porto Alegre, RS, BRAZIL}

\maketitle
\begin{abstract}
We estimate the perturbative nuclear shadowing  in the nuclear structure function $F_2^A(x,Q^2)$, mainly at the kinematic region of HERA using the Glauber-Mueller approach. The  contributions of the quark and gluon sectors to  the nuclear shadowing are estimated.  We predict that the nuclear shadowing corrections are important and that  
 saturation of the ratio $R_1 = F_2^{A}(x,Q^2)/A\,F^p_2(x,Q^2)$  occurs once  the shadowing in the gluon sector is considered.
\end{abstract}


\bigskip

%\begin{multicols}{2}

The future ultrarelativistic heavy ion collider experiments at the BNL Relativistic Heavy Ion Collider (RHIC) and the CERN Large Hadron Collider (LHC) are expected to exhibit new phenomena associated with an ultradense environment that may be created in the central collision region of these reactions \cite{harris}. The main conclusion emerging from the analysis of nucleus-nucleus collisions for RHIC energies and beyond, is that the most of the entropy and transverse energy is presumably produced already during very early times (within the first 2 $fm$ after the nuclear contact) by frequent, mostly inelastic, semihard gluonic collisions involving typical momentum transfers of only a few $GeV$ \cite{eskm}. However, one is still far from a complete and detailed picture, as  reflected by the considerable uncertainty in perturbative QCD predictions for global observables in nucleus-nucleus ($AA$) collisions at collider energies, such as particle multiplicities and transverse energy production \cite{rep}. The inability to extrapolate accurately from the $pp$ ($p\overline{p}$) data to heavy ion $AA$ collisions is due the current poor knowledge about the nuclear and dense medium effects.  One of these effects is the nuclear shadowing. It is of interest in  high energy nucleus-nucleus collisions because it could influence significantly the initial conditions in such reactions. Some processes strongly affected by nuclear shadowing at collider energies include the  minijet production, heavy quarks, their bound states, and dilepton production \cite{rep}. Consequently, the nuclear shadowing is one of the major theoretical issues in modeling the QCD processes in nuclear collisions.

In recent years several experiments have been dedicated to high precision measurements of deep inelastic lepton scattering (DIS) off nuclei. Experiments at CERN and Fermilab focus especially on the region of small values of the Bjorken variable $x = Q^2/2M\nu$, where $Q^2=-q^2$ is the squared four-momentum transfer, $\nu$ the energy transfer and $M$ the nucleon mass.
The data \cite{arneodo,e665}, taken over a wide kinematic range $10^{-5}\,\le\,x\le\,0.1$ and  $0.05\,GeV^2\,\le\,Q^2\le\,100\,GeV^2$, show a systematic reduction of the nuclear structure function $F_2^A(x,Q^2)/A$ with respect to  the free nucleon structure  function $F_2^N(x,Q^2)$. This phenomena is known as {\it the shadowing effect}. The analysis of the shadowing corrections for the nuclear case in deep inelastic scattering (DIS) has been extensively discussed \cite{hera96}. It is motivated by the perspective that in a near future an experimental investigation of the nuclear shadowing at small $x$ and $Q^2 >> 1 \, GeV^2$ using $eA$ scattering could occur at DESY Hadron Electron Ring Accelerator (HERA).

The deep inelastic scattering off a nucleus is usually interpreted in a frame where the nucleus is going very fast. In this case the nuclear shadowing is a result of an overlap in the longitudinal direction of the parton clouds originated from different bound nucleons \cite{qiu}.
Recently, a perturbative approach has been developed to calculate the gluon distribution in a nucleus \cite{ayala1} using perturbative QCD at small $x$.
This approach, known as  Glauber-Mueller approach is formulated in the target rest frame, takes into account the fluctuations of the hard probe.
It  includes the shadowing corrections (SC) due to parton rescatterings  inside the nucleus,
and provides the SC to the nuclear gluon distribution using the solution of the DGLAP evolution equations \cite{dglap} to the nucleon case.
As a result the behavior of related observables ($F_2^A, F_L^A$, ...) at high energies can be calculated.

Our goal in this Letter is to estimate  the perturbative nuclear shadowing  to the nuclear structure function $F_2^A(x,Q^2)$, using the Glauber-Mueller approach proposed in \cite{ayala1} at HERA-A kinematic region.
 We  compute the nuclear shadowing corrections to the structure function
$F_2^A$ associated with the rescatterings of the $q\overline{q}$ pair on the nucleons inside a nucleus using the Glauber-Mueller approach. Our results are dependent of the behavior of the nucleon gluon distribution.  We then consider two possibilities. First we contemplate the unshadowed gluon distribution, which we denote the quark sector contribution for the nuclear shadowing. Secondly  the shadowed gluon distribution, which we denote as the gluon sector contribution for the nuclear shadowing. This contribution is  calculated   using the Glauber-Mueller approach.
Our work is motivated by our recent results  in the analysis of scaling violations of the proton structure function. The ZEUS data for the $F_2^p$ slope presents a turn over which  can be successfully described considering the contribution of both sectors. Here we estimate the contribution  of both sectors  to the ratio  $R_1 = F_2^{A}(x,Q^2)/A\,F^p_2(x,Q^2)$ and we show that the inclusion of the contribution of the gluon sector modify strongly the behavior of this ratio.

Before we calculate the perturbative SC at $F_2^A$ some comments are in order.
As $x \approx Q^2/s$, where $s$ is the squared CM energy, the   data in the region of small $x$ values are for small values of $Q^2$ ($ \le 1\,GeV^2$). In this region the application of the perturbative QCD cannot be justified. The SC in this region are dominated by soft contributions, as can be concluded  considering some approaches \cite{barone} \cite{capella} \cite{kumano} \cite{eskola} that describe these data. For instance, the NMC  experimental results \cite{arneodo} can be described using the DGLAP evolution equations \cite{dglap}
with adjusted initial parton distributions, following refs. \cite{kumano,eskola}. Since the initial parton distributions are associated with  nonperturbative QCD this approach intrinsically includes the nonperturbative contributions to the shadowing corrections.
Therefore, we consider that   the nuclear shadowing corrections observed in the NMC \cite{arneodo} and E665 data \cite{e665} at small $x$ are dominated by soft contributions. The determination of the soft contributions to the nuclear shadowing is not the goal of this Letter.

Let us start from the Glauber-Mueller approach proposed in \cite{ayala1}. In the small $x$ region the gluon distribution governs the behavior of the observables. In  the nucleus rest frame we can consider the interaction between a  virtual colorless hard probe and the nucleus via a gluon pair ($gg$) component of the virtual probe. In the region where 
$x << 1/2mR$ ($R$ is the size of the target), the $gg$ pair 
crosses the target with fixed
transverse distance $r_t$ between the gluons. 
Moreover, at high energies the lifetime of the $gg$ pair  may substantially exceeds the nuclear radius. The cross section for this process is written as \cite{ayala1}
\begin{eqnarray}
\sigma^{g^*A}=  \int_0^1 dz \int \frac{d^2r_t}{\pi} 
 |\Psi_t^{g^*}(Q^2,r_t,x,z)|^2 \sigma^{gg+A}(z,r_t^2)\,\,,
\label{sig1}
\end{eqnarray}
where $g^*$ is the  virtual colorless hard probe with virtuality  $Q^2$, $z$ is the fraction of energy carried by the gluon and $\Psi_t^{g^*}$ is the wave function of the transverse polarized gluon in the virtual probe. Furthermore, $\sigma^{gg+A}(z,r_t^2)$ is the cross section of the interaction of the $gg$ pair with the  nucleus.

To estimate the nuclear shadowing we have to take into account the rescatterings of the gluon pair inside the nucleus. The contributions of the  rescatterings were estimated using the Glauber-Mueller approach  in  ref. \cite{ayala1}. One of the main results is that the $\sigma^{g^*A}$ cross section,  which can be obtained considering the $s$-channel unitarity and the eikonal model, is given by   
\begin{eqnarray}
\sigma^{g^*A}=  \int_0^1 dz \int \frac{d^2r_t}{\pi} 
\int \frac{d^2b_t}{\pi} |\Psi_t^{g^*}(Q^2,r_t,x,z)|^2 \,2\,[1 - e^{-\frac{1}{2}\sigma_N^{gg}(x^{\prime}
,\frac{r_t^2}{4})S(b_t)}]\,\,, 
\label{sig2}
\end{eqnarray}
where $b_t$ is the impact parameter and $\sigma_N^{gg}$ is the cross section of the interaction of the $gg$ pair with the  nucleons inside of the nucleus.
Using the relation $\sigma^{g^*A}(x,Q^2) = (4\pi^2 \alpha_s/Q^2)xG_A(x,Q^2)$ and the expression of the wavefunction $\Psi^{g^*}$ calculated in \cite{mue,ayala1},  the Glauber-Mueller formula for the nuclear gluon  distribution  is obtained  as
\begin{eqnarray}
xG_A(x,Q^2) = \frac{4}{\pi^2} \int_x^1 \frac{dx^{\prime}}{x^{\prime}}
\int_{\frac{4}{Q^2}}^{\infty} \frac{d^2r_t}{\pi r_t^4} \int_0^{\infty}
\frac{d^2b_t}{\pi}\,2\,[1 - e^{-\frac{1}{2}\sigma_N^{gg}(x^{\prime}
,\frac{r_t^2}{4})S(b_t)}]\,\,.
\label{gluon}
\end{eqnarray}
The equation (\ref{gluon}) allows to estimate the value of the nuclear shadowing corrections in the nuclear gluon distribution.

In \cite{plb} the authors demonstrate that $ \sigma_N^{gg} = (3 \alpha_s(\frac{4}{r_t^2})/4)\,\pi^2\,r_t^2\,
 xG_N(x,\frac{4}{r_t^2})$, where  $xG_N(x,Q^2)$ is the nucleon gluon distribution.  Therefore, the behavior of the cross section (\ref{sig2}) and of the nuclear gluon distribution (\ref{gluon}) in the small-$x$ region are driven by the behavior of the nucleon gluon distribution in this region.
The Glauber-Mueller formula (\ref{gluon}) 
 gives us the possibility to  calculate the SC using the solution of the DGLAP evolution equation. In this Letter, we use the GRV parameterization \cite{grv95} as a solution of the DGLAP evolution equation. It  describes all available nucleon experimental data quite well for $Q^2 \ge 1 \, GeV^2$ \cite{h1}. It should also be stressed here that we disregard how much of SC have been taken into account in this parameterization in the form of the initial distributions. The possible SC in the initial conditions are nonperturbative contributions to shadowing.

The use of the Gaussian parameterization for
the   profile function $S(b_t) = (A/\pi R^2) exp(-b_t^2/R^2)$ simplifies our calculations. Then, doing the integral over $b_t$,  the  master equation is obtained  as
\begin{eqnarray}
xG_A(x,Q^2) = \frac{2R_A^2}{\pi^2}\int_x^1 \frac{dx^{\prime}}{x^{\prime}}
\int_{\frac{1}{Q^2}}^{\frac{1}{Q_0^2}} \frac{d^2r_t}{\pi r_t^4} \{C + ln(\kappa_G(x^{\prime}, r_t^2)) + E_1(\kappa_G(x^{\prime}, r_t^2))\}  
\label{master}
\end{eqnarray} 
where $C$ is the Euler constant,  $E_1$ is the exponential function, the function  $\kappa_G(x, r_t^2) = (3 \alpha_s\,A/2R_A^2)\,\pi\,r_t^2\,
 xG_N(x,\frac{1}{r_t^2})$, 
 $A$ is the number of nucleons in a nucleus and $R_A^2$ is the mean nuclear radius. If equation (\ref{master}) is expanded for small $\kappa_G$, the first term (Born term) will correspond to the usual DGLAP equation in the small $x$ region, while the other terms will take into account the shadowing corrections.

The master formula  (\ref{master}) is obtained in the double logarithmic approximation (DLA) \cite{ayala1}. As shown in \cite{ayala1} the DLA does not work quite well in the accessible kinematic region ($Q^2 > 1\,GeV^2$ and $x > 10^{-4}$).  A more realistic approach must be considered to calculate the nuclear gluon distribution $xG_A$, which implies the subtraction of the Born term of   (\ref{master}) and the sum of the GRV parameterization  \cite{ayala1}. This procedure gives
\begin{eqnarray}
xG_A(x,Q^2)  =  xG_A^{master} \,[\mbox{eq. (\ref{master})}] + AxG_N^{GRV}(x,Q^2)  -  A\frac{\alpha_s N_c}{\pi} 
\int_x^1 \int_{Q_0^2}^{Q^2}  \frac{dx^{\prime}}{x^{\prime}}  \frac{dQ^{\prime 2}}{Q^{\prime 2}}\,x^{\prime}G_N^{GRV}(x^{\prime},Q^{\prime 2})\,\,.
\label{gluon2}
\end{eqnarray}
The above equation implies $AxG_N^{GRV}(x,Q_0^2)$ as the initial condition for the gluon distribution and gives  $AxG_N^{GRV}(x,Q^2)$ as the first term of the expansion with respect to $\kappa_G$. Therefore, this equation is an attempt to include the full expression for the anomalous dimension for the scattering off each nucleon, while the use of  the DLA  takes into account all SC. In \cite{ayala1} this procedure was applied to obtain the shadowing corrections to the nuclear gluon distribution, demonstrating that the suppression due to the shadowing corrections increases with $ln \, 1/x$ and is much bigger than  the nucleon case. For  Calcium ($A=40$) the suppression varies from $4\%$ for $ln \, 1/x = 3$ to $25\%$ for $ln \, 1/x = 10$.

The expression (\ref{gluon2}) estimates the SC to the gluon distribution using the Glauber-Mueller approach. The modification in the nuclear gluon distribution  represents  the contribution of the gluon sector for the nuclear shadowing.

%In this letter we obtain the SC to the nuclear structure function $F_2^A(x,Q^2)$. 

The Glauber-Mueller approach can be extended to the nuclear structure function considering that in the rest frame of the target the virtual photon decays  into a quark-antiquark ($q\overline{q}$) pair long before the 
interaction with the target. The $q\overline{q}$ pair subsequently interacts 
with the target.  In the small $x$ region, where 
$x << 1/2mR$, the $q\overline{q}$ pair 
crosses the target with fixed
transverse distance $r_t$ between the quark and the antiquark. Considering the $s$-channel unitarity and the eikonal model and following the same steps used in the case of the nuclear gluon distribution,  the nuclear structure function  can be written as
\begin{eqnarray}
F_2^A(x,Q^2) =  \frac{2R_A^2}{3\pi^2} \sum_1^{n_f} \epsilon_i^2 \int_{\frac{1}{Q^2}}^{\frac{1}{Q_0^2}} \frac{d^2r_t}{\pi r_t^4} \{C + ln(\kappa_q(x, r_t^2)) + E_1(\kappa_q(x, r_t^2))\}\,\,,
\label{diseik}
\end{eqnarray}
where $\kappa_q = 4/9 \, \kappa_G$. Similarly as made in the case of the nuclear gluon distribution, to obtain a more realistic approach the Born term should be subtracted and the GRV parameterization should be added. Therefore the nuclear structure function is given by
\begin{eqnarray}
F_2^A(x,Q^2) = F_2^A(x,Q^2) [\mbox{eq. (\ref{diseik})}] - F_2^A(x,Q^2)[\mbox{Born}] + A\,F_2(x,Q^2) [\mbox{GRV}] \,\,,
\label{f2ab}
\end{eqnarray}
where $ F_2^A(x,Q^2)[\mbox{Born}]$ is the first term in the expansion in $\kappa_q$ of the equation (\ref{diseik}), and 
$F_2(x,Q^2) [\mbox{GRV}] = \sum_{u,d,s} \epsilon_q^2 \,[xq(x,Q^2) + x\overline{q}(x,Q^2)] + F_2^c(x,Q^2)$
is calculated using the GRV parameterization.
The charm component of the structure function is calculated  considering the charm production via boson-gluon fusion \cite{grv95}. In this Letter we assume $m_c = 1.5\,GeV$.

The expression (\ref{f2ab}) estimates the SC to the nuclear  structure function  using the Glauber-Mueller approach. We see that the behavior of $F_2^A$ is directly associated with the behavior of the gluon distribution used as input, which is usually  described by a parameterization of the parton distributions \cite{grv95}. In this case the shadowing in the gluon distribution is not included explicitly. We denote  the quark  sector contribution  for the nuclear shadowing the effect in the nuclear structure function obtained using the Glauber-Mueller approach and a usual parameterization (unshadowed)  as input.

The total modification in the structure function associated with the nuclear shadowing is the sum of the shadowing corrections in  both sectors.
The main difference between quark and gluon sectors  stems from the much larger cross section $\sigma_N^{gg} = (9/4) \sigma_N^{q\overline{q}}$, {\it i. e.}  $\kappa_G = (9/4) \kappa_q$, which in turn leads to a much larger gluon shadowing. Therefore,  at small values of $x$ the contribution of the gluon sector cannot be disregarded in  a reliable calculus of the nuclear structure function. In a general case, the shadowing corrections for $F_2^A$ should be estimated considering also these corrections for the gluon distribution, {\it i.e.} in the quark and gluon sectors. Recently, we propose  a  procedure to estimate the SC in both sectors  to describe the turn over in the $F_2$ slope observed in the ZEUS data \cite{zeusdf2}.
Using the procedure here, we estimate the nuclear structure function considering both sectors (quark+gluon sectors) of NS using the solution of (\ref{gluon2}) for the nucleon case ($ A = 1$) as an input in the expression  (\ref{f2ab}).

Using  the expression (\ref{f2ab}) we can estimate the nuclear shadowing corrections. We assume that the mean radius of the nucleus $R_A^2$ is equal to $R_A^2 = 2/5\,R_{WS}^2$, where $R_{WS}$ is the size of the nucleus in the Woods-Saxon parameterization. We choose  $R_{WS} = r_0 A^{\frac{1}{3}}$, with $r_0=1.3 \, fm$.
We calculate the ratio
\begin{eqnarray}
R_1 = \frac{F_2^{A}(x,Q^2)}{A\,F^p_2(x,Q^2)}\,\,,
\label{ratio}
\end{eqnarray}
where $F_2^{A}(x,Q^2)$ is the  structure function for a nucleus with $A$ nucleons and $F^p_2(x,Q^2)$ is the proton  structure function.

\begin{figure}[t]
\begin{tabular}{c c}
\psfig{file=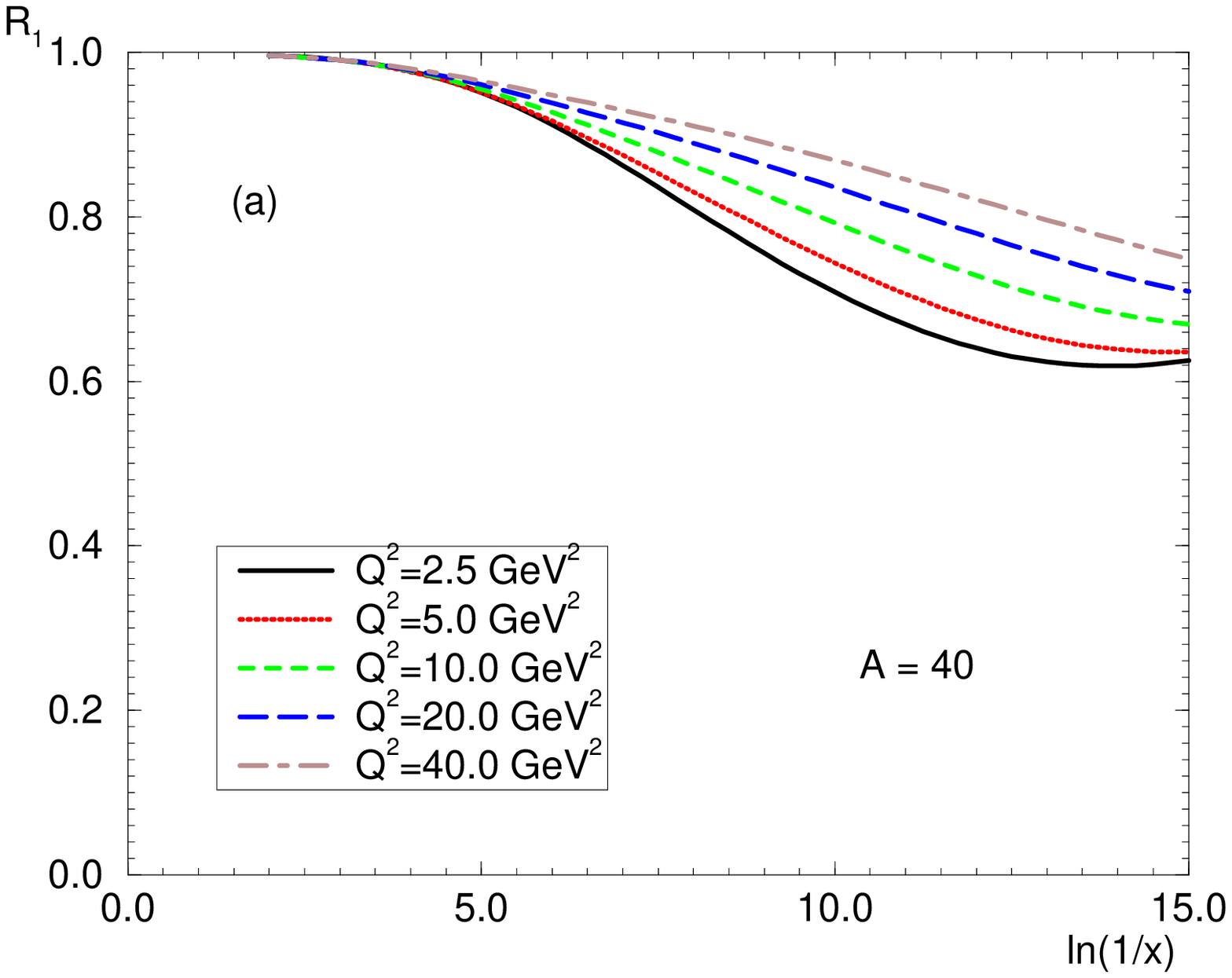,width=80mm} & \psfig{file=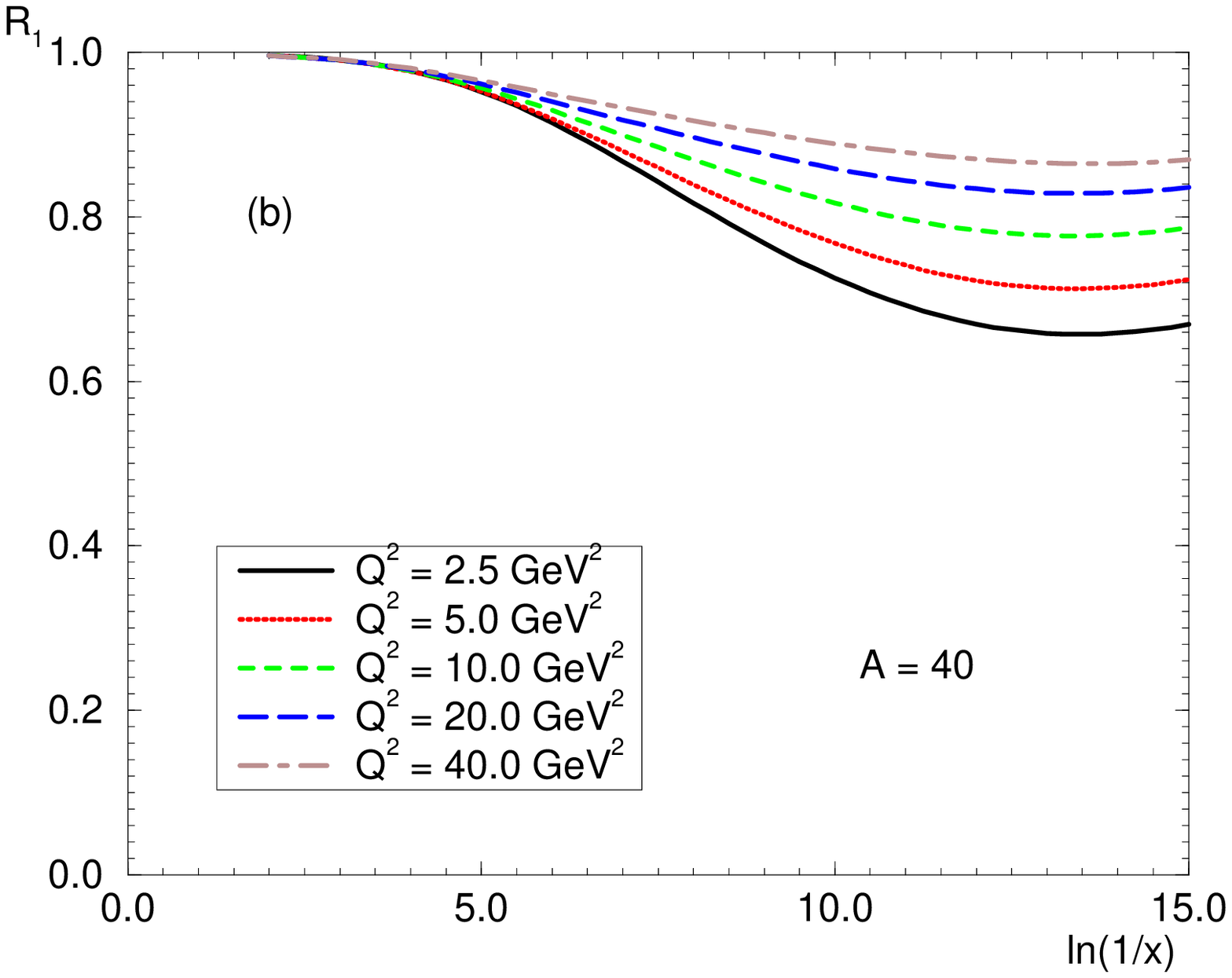,width=80mm} \end{tabular}
\caption{ The ratio  $R_1 = F_2^{A}(x,Q^2)/A\,F^p_2(x,Q^2)$  as a function of $ln\,1/x$ at different virtualities and different sectors: (a) quark sector  and (b) quark+gluon sector. In the Calcium case $A=40$.}
\label{fig1}
\end{figure}

In figure \ref{fig1} we present our results for the ratio  $R_1$ as a function of $ln\,1/x$ at different virtualities and considering the different sectors for the Calcium case  ($A=40$). In Fig. \ref{fig1}(a) we present the results of the NS for the ratio considering only the quark sector, while in Fig. \ref{fig1}(b) we present the results of NS corrections considering the quark+gluon sectors. We can see that in both cases the perturbative nuclear shadowing is an important correction to the behavior of the nuclear structure function. The suppression varies from $1\%$ for $ln \, 1/x = 3$ to $20\%$ for $ln \, 1/x = 10$. Therefore our result demonstrates that the suppression due to shadowing corrections is proportionally smaller in the nuclear structure function than in the nuclear gluon distribution. This result is in disagreement with the result obtained in  ref. \cite{capella} which uses the Gribov formula and a Regge parameterization for the diffractive structure function.  We believe that this disagreement occurs since that approach is dominated by soft physics. 
Furthermore, we can see that when only the contribution of the quark sector is considered  no saturation in the ratio $R_1$ is observed, which agrees with the result obtained in \cite{ina}. However, if the contribution of the gluon sector is considered, {\it i.e.} if a shadowed gluon distribution is used,  the ratio saturates in the HERA-A kinematical region.  This is a new result, since that in general the saturation  in  ratio $R_1$ has been characterized as   essentially  a nonperturbative effect \cite{ina}.  Therefore,  a  saturation in the perturbative region ($Q^2 > 1\, GeV^2$) evidentiates the large shadowing corrections in the gluon sector.

%no decrease of shadowing will be observed at HERA-A. Our result disagrees with %the result obtained in %\cite{povh} which predicts  the vanishing of the nuclear shadowing for $x\rightarrow 0$ at fixed $Q^2$. %We believe that this %disagreement occurs since this approach is dominated by soft physics. Therefore, we %consider that the observed saturation of the shadowing ratio \cite{e665} at small $x$ is thus a %non-perturbative, low-$Q^2$ feature.

\begin{figure}
\centerline{\psfig{file=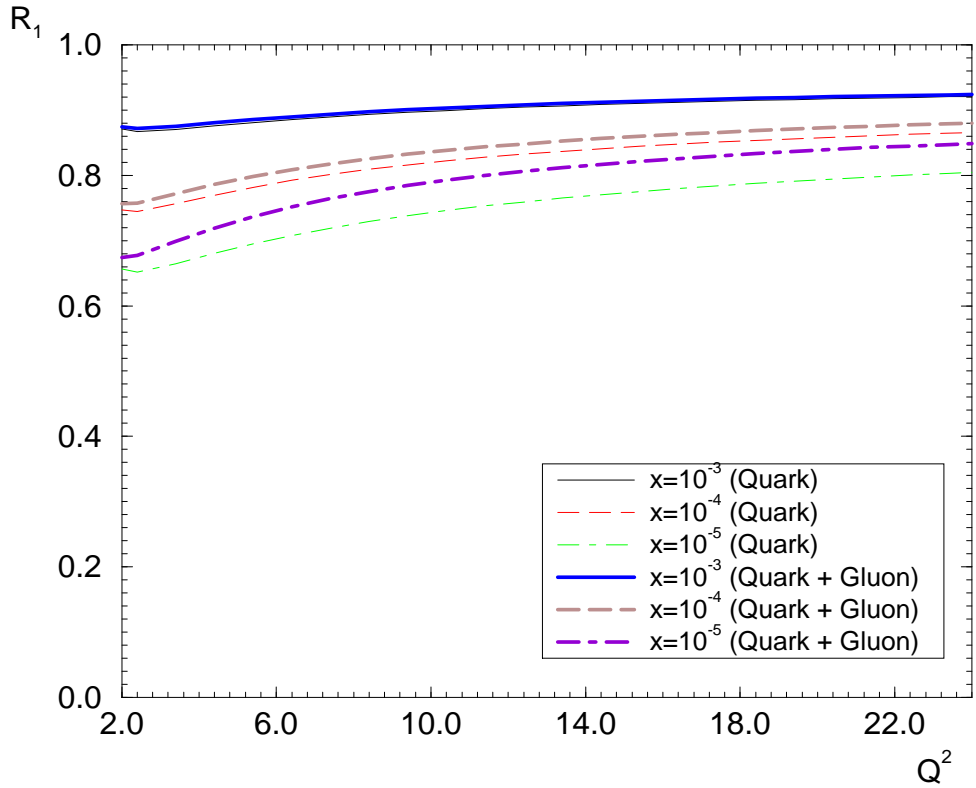,width=150mm}}
\caption{The ratio $R_1 = F_2^{A}(x,Q^2)/A\,F^p_2(x,Q^2)$  as a function of $Q^2$ at different values of $x$ and different sectors (quark and quark+gluon).  In the Calcium case $A=40$.}
\label{fig2}
\end{figure}

In figure \ref{fig2} we present our results for the ratio $R_1$ as a function of $Q^2$ at different values of $x$ and different sectors for the Calcium case  ($A=40$). As the behavior of the ratio $R_1$ in the region $Q^2 \le 2.5\,GeV^2$ is strongly dependent of the initial conditions \cite{ina}, we present only our  predictions in the region of larger $Q^2$.  We can see that the ratio is dependent on $Q^2$ in the region $Q^2 \le 10\,GeV^2$ and that  at large $Q^2$ the nuclear shadowing corrections diminish but do not vanish, {\it i.e.} it is not a higher twist effect. In addiction, for $x\ge 10^{-3}$ the curves associated with  the quark and quark+gluon sector coincides, which represents that gluon sector do not contribute in this kinematical region. The dependence of the ratio $R_1$ on $Q^2$ occurs mainly at small values of $x$, but the general behavior of the curve do not depend on the sector considered.

\begin{figure}
\centerline{\psfig{file=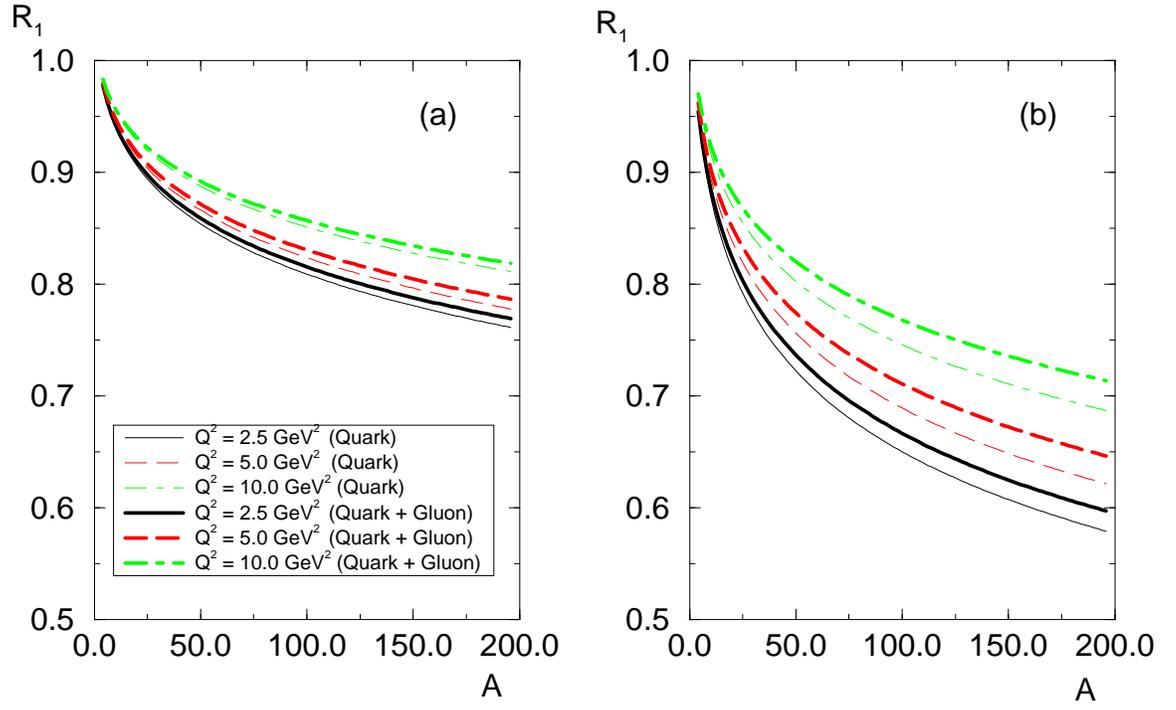,width=150mm}}
\caption{The ratio  $R_1 = F_2^{A}(x,Q^2)/A\,F^p_2(x,Q^2)$   as a function of $A$ at different values of $Q^2$, different sectors (quark and quark+gluon) and (a) $x=10^{-3}$, (b)  $x=10^{-4}$.}
\label{fig3}
\end{figure}

In figure \ref{fig3} we present our results of the ratio $R_1$ as a function of $A$ at different values of $Q^2$ and $x$ for the different sectors. We can see that the nuclear shadowing corrections increases with the growth of $A$ and with the decrease of $x$. The nuclear shadowing corrections for the ratio $R_1$ are  larger  for smaller $Q^2$ values and smaller if the gluon sector is considered.

In summary,  the perturbative nuclear shadowing corrections for HERA-A are analyzed in this Letter in the context of the Glauber-Mueller approach. This approach  predicts important results for the  perturbative nuclear shadowing: the nuclear shadowing corrections are very large in the HERA-A kinematical region; 
 saturation of the nuclear shadowing corrections will occur in this kinematical region if the contribution of the gluon shadowing is considered; the shadowing in the nuclear gluon distribution is  larger than in the nuclear structure function.
We expect that these results motivates the running of nucleus at HERA in the future, since  the precise determination of nuclear shadowing corrections is fundamental to estimate the cross sections of the processes with nucleus which will be studied in the future accelerators RHIC and LHC.

\section*{Acknowledgments}

MBGD acknowledges F. Halzen for useful discussions.
This work was partially financed by CNPq and by Programa de Apoio a N\'ucleos de Excel\^encia (PRONEX), BRAZIL.

\newpage

%\begin{figure}[t]
%\begin{tabular}{c c}
%\psfig{file=fig3.eps,width=70mm} & \psfig{file=fig4.eps,width=70mm} %\end{tabular}
%\caption{}
%\label{fig3}
%\end{figure}

%\end{multicols}

\end{document}